\documentclass[a4paper,11pt]{article}
\usepackage[export]{adjustbox}
\usepackage[english]{babel}
\usepackage[utf8]{inputenc}
\usepackage{pos}
\usepackage{amsmath}
\usepackage{color,graphicx}
\usepackage{bm,bbm,braket}
\usepackage{comment}
\usepackage{enumerate}
\usepackage{hyperref}
\usepackage{multirow}
\usepackage{mathtools}
\usepackage{mathrsfs}
\usepackage{mciteplus}
\usepackage{slashed}
\usepackage{soul}
\usepackage{subfigure}
\usepackage{wrapfig}
\usepackage{xspace}

\newcommand{\p}{\bm{p}}
\newcommand{\q}{\bm{q}}

\newcommand{\2}{\bm{2}}
\newcommand{\3}{\bm{3}}

\newcommand{\n}{\bm{n}}
\newcommand{\m}{\bm{m}}

\newcommand{\Ac}{\mathcal{A}}
\newcommand{\wAc}{\widetilde{\mathcal{A}}}
\newcommand{\Bc}{\mathcal{B}}

\newcommand{\Dc}{\mathcal{D}}

\newcommand{\Fc}{\mathcal{F}}
\newcommand{\Gc}{\mathcal{G}}
\newcommand{\Hc}{\mathcal{H}}
\newcommand{\Ic}{\mathcal{I}}

\newcommand{\Kc}{\mathcal{K}}

\newcommand{\Mc}{\mathcal{M}}

\newcommand{\Rc}{\mathcal{R}}

\DeclareMathOperator{\re}{Re}
\DeclareMathOperator{\im}{Im}

\newcommand{\bcol}{\left[ \begin{array}{c}}
\newcommand{\ecol}{\end{array} \right]}
\newcommand{\beq}{\begin{eqnarray}}
\newcommand{\eeq}{\end{eqnarray}}

\newcommand{\kev}{\ensuremath{{\mathrm{\,ke\kern -0.1em V}}}\xspace}
\newcommand{\mev}{\ensuremath{{\mathrm{\,Me\kern -0.1em V}}}\xspace}
\newcommand{\gev}{\ensuremath{{\mathrm{\,Ge\kern -0.1em V}}}\xspace}
\newcommand{\tev}{\ensuremath{{\mathrm{\,Te\kern -0.1em V}}}\xspace}

\usepackage{mfirstuc} 
\newcommand{\addReviewer}[2]{
  \expandafter\newcommand\csname #1\endcsname[1]{{\bf \color{#2} \capitalisewords{#1}:\,##1}}
  \expandafter\newcommand\csname #1cor\endcsname[2]{{\color{#2} \capitalisewords{#1}:\,\st{##1}{\bf ##2}}}
  \expandafter\newcommand\csname #1color\endcsname{#2}
}

\usepackage{soul,color}
\definecolor{chromeyellow}{rgb}{1.0, 0.65, 0.0}
\definecolor{DodgeBlue}{rgb}{0.118, 0.565,1.000}
\definecolor{asparagus}{rgb}{0.53, 0.66, 0.42}
\definecolor{cardinal}{rgb}{0.77, 0.12, 0.23}
\definecolor{cadmiumgreen}{rgb}{0.0, 0.42, 0.24}
\definecolor{applegreen}{rgb}{0.55, 0.71, 0.0}

\addReviewer{wrong}{red}
\addReviewer{good}{green}
\addReviewer{sebastian}{cardinal}

\title{Infinite volume, three-body scattering formalisms in the presence of bound states}

\author*[a,b]{Sebastian M. Dawid}

\affiliation[a]{Physics  Department, Indiana  University, Bloomington,  IN  47405,  USA}

\affiliation[b]{Center for  Exploration  of  Energy  and Matter, Indiana  University, Bloomington,  IN  47403,  USA}

\emailAdd{sdawid@iu.edu}

\abstract{Strong interactions produce a rich spectrum of resonances that decay into three or more hadrons. Understanding their phenomenology requires a theoretical framework to extract parameters from experimental data and Lattice QCD simulations of hadron scattering. Two classes of relativistic three-body approaches are currently being pursued: the EFT-based and unitarity-based one. We consider a model of relativistic three-body scattering with an $S$-wave bound state in the two-body sub-channel using both formalisms. We present and discuss numerical solutions for the multi-hadron scattering amplitudes in different kinematical regions, obtained from integral equations of the EFT-based approach. The connection of our work to the ongoing program of computing the three-body spectrum from the lattice is highlighted. Finally, we show how to generalize the unitarity-based framework to include all relevant open channels, discuss the nonphysical singularities near the physical region, and show how to eliminate them in a simple case.}

\FullConference{%
 The 38th International Symposium on Lattice Field Theory, LATTICE2021
  26th-30th July, 2021
  Zoom/Gather@Massachusetts Institute of Technology
}

\begin{document}
\maketitle


\section{Introduction}

It is well-established experimentally that many resonances couple strongly to three- or more particle channels
\citep{KETZER2020103755,Olsen:2017bmm}. 
Some of the most puzzling particles like Roper resonance $N^{*}(1440)$ or exotic $\chi_{c1}(3872)$ have significant three-particle decay modes \citep{ParticleDataGroup:2020ssz}. A general theoretical framework describing the scattering of three hadrons would provide a wider avenue for systematic phenomenological studies of those states. Moreover, such a framework is required to establish their properties directly from QCD, translating the lattice results for the three-particle spectra into infinite-volume scattering information \citep{Luscher:1986pf,Luscher:1990ux,Briceno:2017max}. In recent years, two relativistic on-shell $3\! \to \! 3$ scattering formalisms have been developed and applied to a range of physical problems: \textbf{(a)} the relativistic EFT (RFET)~\citep{Hansen:2014eka,Hansen:2015zga,Briceno:2017tce,Briceno:2018aml}, and \textbf{(b)} the $S$-matrix unitarity approach~\citep{Mai:2017vot,Mai:2017bge,Doring:2018xxx,Jackura:2018xnx}. In both formalisms, the lattice output, generically called the three-body $K$-matrix, enters into a set of integral equations. Their solution yields an on-shell three-particle scattering amplitude \citep{Hansen:2019nir,Mai:2021lwb}, which has to be continued to the complex energies to identify resonances. Due to the multi-variable nature of three-body scattering, these two last steps pose a challenge. First, the integral equations require a careful numerical solution procedure to arrive at correct results. Additionally, before performing analytical continuation, one needs to understand the analytic properties of the amplitudes to disentangle kinematic or unphysical singularities from the physical ones.

\begin{figure}[t!]
    \centering
    \includegraphics[ width=0.55\columnwidth, trim= 4 4 4 4,clip]{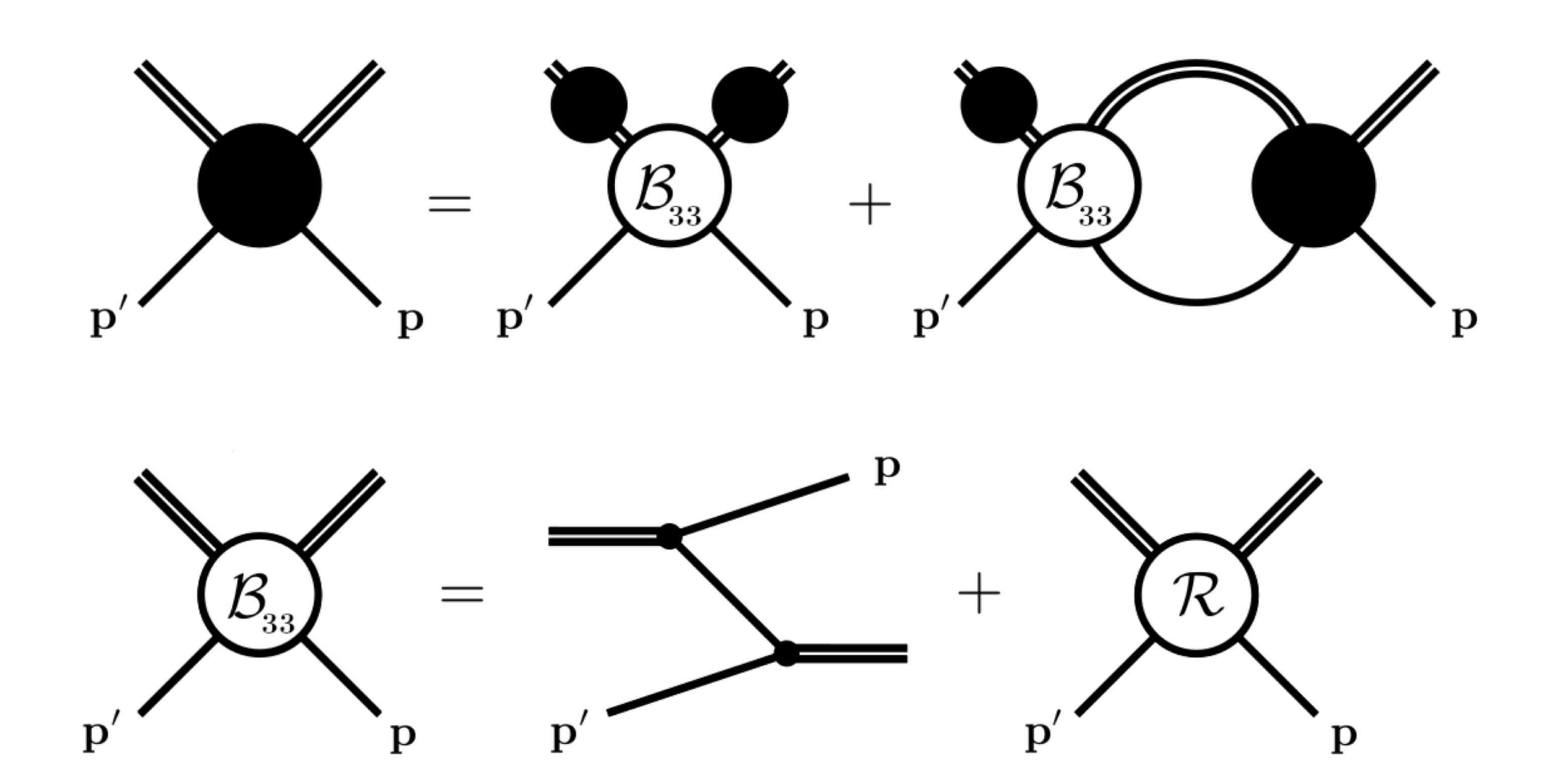}
    \put(-250,87){\colorbox{white}{(a)}}
    \put(-250,25){\colorbox{white}{(b)}}
    \caption{
    Diagrammatic representation of (a) $\Ac_{33}$, as given by Eq.~\eqref{eq:b-matrix-param}, and (b) the $B$-matrix kernel. A single external line represents a spectator, while a double external line a pair. A solid circle with both external pairs and spectators is the three-body connected amplitude $\Ac_{33,\p'\p}$, and a solid circle only with external pairs is the two-body amplitude $\Fc_{\p}$.}
    \label{fig:b-matrix}
    \end{figure}

To resolve the first of those problems, in Ref.~\citep{Jackura:2020bsk} we established a systematically improvable method for numerically solving relativistic three-body integral equations of the type depicted diagrammatically in Fig.~\ref{fig:b-matrix}. We test it by solving a three-body problem of scalar particles with an $S$-wave two-body bound state—a toy model for the nucleon-deuteron interaction. The problem is numerically demanding due to the pole singularity appearing in the integration range. Here, for clarity of presentation, we describe the simplest employed numerical method, i.e., basic uniform discretization. We briefly discuss systematic tests utilized to check the quality of our solutions, present example results, and compare them to the existing studies.

Addressing the second of the above-mentioned problems, in Ref.~\citep{Dawid:2020uhn} we studied the problem of the analytical structure of the three-body amplitudes. Using the ``minimal'' $B$-matrix formalism, we analyze the same system, i.e., scattering of three spinless particles, in which a bound state in the two-body subchannel is formed.
In this approach, the intermediate states are on-shell and have physical energies. We find that this generates non-physical analytic properties of the amplitude. In particular, spurious singularities appear arbitrarily close to the three-body threshold, hindering the study of genuine three-body effects.
We show how to eliminate non-physical singularities by employing Chew--Mandelstam-like dispersion procedure when the contact-interaction approximation is assumed. We also generalize the formalism to include all relevant open channels, and corresponding scattering amplitudes. Finally, we show how the dispersion relations ensure they satisfy the unitarity constraint between the bound-state--particle and three-body thresholds.

\section{Three-body integral equations}
\label{sec:Integral-equations}

\begin{figure}[t!]
    \centering
    \includegraphics[scale=0.3]{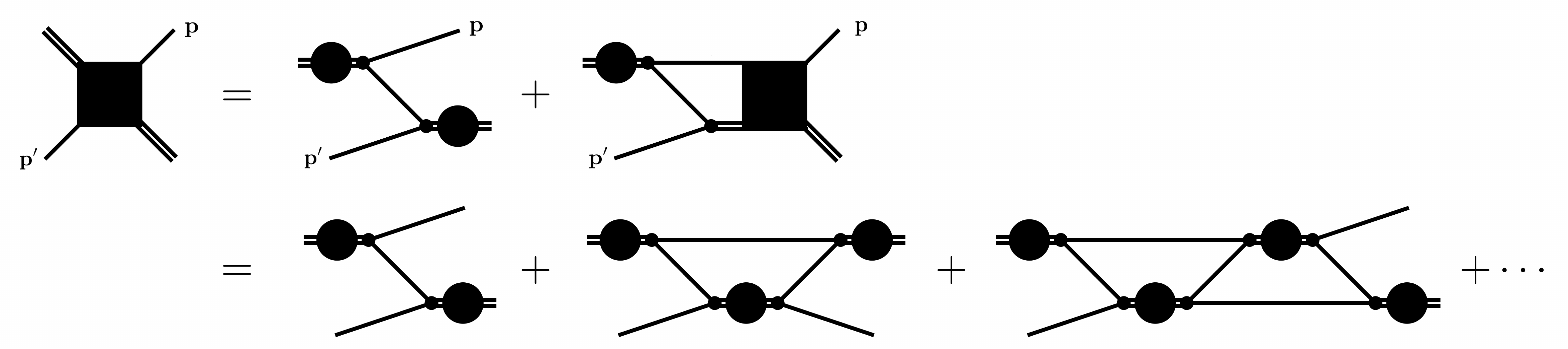}
    \caption{Diagrammatic representation of the three-body ``ladder'' integral equation. A rectangle with pair-spectator external legs is the ladder three-body amplitude.}
    \label{fig:ladder}
\end{figure}

Firstly, we summarize the relevant aspects of the REFT and the $B$-matrix formalisms, adapting the notation of Refs.~\citep{Jackura:2019bmu,Dawid:2020uhn}. We consider an elastic scattering process of three indistinguishable spinless particles of mass $m$. The unsymmetrized partial-wave projected three-body amplitude $\Mc_{33,\p'\p}$ is separated into a connected and disconnected part, $\Ac_{33,\p'\p}$ and $\Fc_{\p}\, \delta_{\p'\p}$,
    \beq
    \label{eq:dis-con-decomp}
    \Mc_{33,\p'\p} = \Ac_{33,\p'\p} + \Fc_{\p} \, \delta_{\p'\p} \, .
    \eeq
Equation~\eqref{eq:dis-con-decomp} is written in the so-called $(\p \ell m_{\ell})$ basis in which amplitudes are treated as infinite-dimensional matrices in the angular momentum space. Here $\bm{p}$ is the momentum of one of the particles in the initial state. We call this particle the \emph{spectator}, whereas the other two form a \emph{pair}, corresponding to the given spectator. The particles in a pair are projected to definite angular momentum $(\ell,m_\ell)$ in their center-of-momentum (CM) frame. In addition to those variables, the amplitude depends on the analogous primed variables for the final state, on the total CM invariant mass $s$ of the three-particle system, and the total angular momentum $J$. In the $B$-matrix approach the connected part of the amplitude is given by the integral equation,
    \beq
    \label{eq:b-matrix-param}
    \Ac_{33,\p' \p} = \Fc_{\p'} \, \Bc_{33,\p' \p} \, \Fc_{\p} + \int_{\bm{k}} \Fc_{\p'} \, \Bc_{33,\p' \bm{k}} \, \Ac_{33,\bm{k} \p} \, ,
    \eeq
as illustrated in Fig.~\ref{fig:b-matrix}. Here, $\Fc_{\p}$ is the $\2 \to \2$ partial wave amplitude that describes interactions of particles in the pair. The $B$-matrix kernel is given by the sum of two terms, $ \Bc_{33,\p'\p} = \Gc_{\p'\p} + \Rc_{\p'\p} $. The matrix $\Gc_{\p'\p}$ governs the long-range interaction due to one-particle exchange (OPE) between the pair and spectator and $\Rc_{\p' \p}$ is a real matrix representing all short-range interactions. The OPE amplitude is specified by the $S$-matrix unitarity, while the $R$-matrix is unconstrained. Assuming $\Rc_{\p'\p}=0$ leads to the solution driven exclusively by the exchanges between $\2\to\2$ sub-processes, see Fig.~\ref{fig:ladder}, and defines the \emph{ladder} amplitude $\Dc_{\p' \p}$, given by,
    \beq
    \label{eq:ladder-1}
    \Dc_{\p' \p} = \Fc_{\p'} \, \Gc_{\p' \p} \, \Fc_{\p} + \int_{\bm{k}} \Fc_{\p'} \, \Gc_{\p' \bm{k}} \, \Dc_{\bm{k} \p} \, .
    \eeq
In the REFT formulation, as described in Ref.~\cite{Hansen:2015zga}, the connected part of the unsymmetrized three-body scattering amplitude is given by the sum of the ladder and short-range amplitude $\Mc^{(u,u)}_{\text{df},3}$. It is obtained from an additional double-integral equation, driven by the three-body $K$ matrix called $\Kc_{\text{df},3}$, representing short-distance three-particle interactions. It is the analog of the $\Rc$ matrix of the $B$-matrix formalism. In the recent lattice studies \citep{Hansen:2020otl,Brett:2021wyd,Mai:2021nul}, both the $\Rc$ and $\Kc_{\text{df},3}$ have been determined for the realistic systems of three pions.

Finally, the integration in the above equations is defined as,
    \beq
    \label{eq:integration}
    \int_{\bm{k}} & \equiv & \int \frac{d \Omega_{\bm{k}}}{4 \pi}\!\! \int\limits_0^{k_{\text{max}}} \!\! \frac{d k \ k^2}{2\pi^2 \omega_{k} } = 
    \int \! \frac{d \Omega_{\bm{k}}}{4 \pi} \!\! \int\limits_{\sigma_{\text{min}}}^{(\sqrt{s}-m)^2} \!\! \frac{d \sigma_{k}}{2 \pi} \, \tau(s, \sigma_{k}) \, .
    \eeq
Here $\bm{k}$ is the intermediate spectator momentum and $\sigma_{k} = (\sqrt{s} - \omega_k)^2 - \bm{k}^2$ is the intermediate pair invariant mass squared and $\omega_k = \sqrt{\bm{k}^2 + m^2}$. Function $\tau(s, \sigma_{\bm{k}}) = \lambda^{1/2}(s, \sigma_{k}, m^2) / 8 \pi s$ is the three-body phase space factor, where $\lambda(x,y,z)$ is the K\"all\'en triangle function. The ``minimal'' $B$-matrix parametrization is defined by $\sigma_{\text{min}} = 4m^2$. That provides a clear distinction between the long-range and short-range effects in the formation of resonances, with the OPE amplitude giving a probability for an exchange of a real, on-shell particle. On the other hand, the REFT formalism takes $\sigma_{\text{min}}=0$ and implements an additional smooth cutoff in the definition of OPE, which has an advantage of pushing left-hand singularities of the amplitude away from the physical energy region.

\section{Ladder equation study}

In Ref.~\citep{Jackura:2020bsk} we develop numerical techniques for solving the ladder equation. For simplicity, we assume that the two-body subsystem contains only contributions from the $\ell = 0$ partial wave, and consider the total angular momentum $J = 0$. Furthermore, we work with the amputated amplitude $d$, defined through relation $\Dc_{\p' \bm{p}} = \Fc_{\p'} \, d_{\p \p} \, \Fc_{\p} $. The ladder equation takes the form\footnote{It should be noted that we use a slightly different notation than Ref.~\citep{Jackura:2020bsk}. We omit $(u,u)$ superscripts and $S$ subscripts, and extract the minus sign from the definition of OPE. We use $\Fc$ for $\Mc_2$, $\sigma_k$ for $s_{2k}$, and $s$ for $s_3$ or $E^2$.},
    \begin{align}
    \label{eq:d_Sproj}
    d(p', p) 
    &=
    G(p', p)   
    + \int_0^\infty \! \frac{\text{d} k \, k^2}{ (2\pi)^2 \, \omega_{k}} \, G(p', k) \, \Fc( k) \, d(k, p) \, .
    \end{align}
The $S$-wave OPE is,
    \begin{align}
    \label{eq:Gs_proj}
    G(p', p) = \frac{H(p',p)}{4p'p} \, \log\left( \frac{\alpha(p',p) - 2p'p + i\epsilon}{\alpha(p',p) + 2p'p + i\epsilon} \right) \, ,
    \end{align}
where $\alpha(p',p) = (\sqrt{s} - \omega_{p'} - \omega_p)^2 - \p^2 - \p'^2 - m^2$ and $H(p',p)$ is the smooth cutoff function defined in Ref.~\citep{Hansen:2015zga}. We consider the leading order effective range expansion for $\Fc$,
    \begin{align}
    \label{eq:ERE}
    \Fc( k) = \frac{16 \pi \sqrt{\sigma_k}  }{-1/a - i \sqrt{\sigma_k /4-m^2} } \,,
    \end{align}
where $a$ is the scattering length. Is has a pole on the real $\sigma_{k}$ axis, which we call $\sigma_b = 4\left(m^2-1/a^2 \right)$, and the residue at the pole $g=8\sqrt{ 2 \pi \sqrt{\sigma_b}/a }$. The relative bound-state--spectator momentum $q$ corresponding to the bound state pole is $q = \lambda^{1/2}(s,s_b,m^2)/2\sqrt{s}$.

To simplify the discussion of the results, we label the spectator as ``$\varphi$'' and the bound state as ``$b$'', and define two-body, bound-state--spectator amplitude $\Mc_{\varphi b}$. It is obtained by continuing the initial and final two-particle subsystems to the bound state pole,
    \begin{align}
    \label{eq:limit}
    \Mc_{\varphi b}(s) =
    \lim_{\sigma_p,\sigma_{p'} \to \sigma_b}
    g^2  \, d( p', p) \, .
    \end{align}
Above the $\varphi b$ threshold $s_{\varphi b} = (\sqrt{\sigma_b} + m)^2$ and below the three-particle threshold $s_{3\varphi} = (3m)^2$, the amplitude satisfies the two-body $S$-matrix unitarity. Therefore, it can be written using the usual $K$-matrix parametrization,
    \begin{align}
    \label{eq:2p1.Kmat}
    {\Mc}_{\varphi b} (s) &= \frac{1}{ \Kc_{\varphi b}^{-1}(s) - i  \rho_{\varphi b}(s) } \, ,
    \end{align}
where $\Kc_{\varphi b}^{-1}(s)$ is real below the $3\varphi$ threshold and $\rho_{\varphi b}(s) = \tau(s,\sigma_b)/2$ is the two-body phase space between the bound state and the spectator. It is worth noting that in our approach one can additionally access the unsymmetrized bound-state breakup amplitude $\varphi b \to 3\varphi$. It is obtained from the limiting procedure in which only the initial pair invariant mass is fixed to the bound-state mass, namely, $\Mc_{\varphi b,3\varphi}( p',s)
= g \, \Fc(p') \lim_{\sigma_{p}\to \sigma_b} \, d( p',  p)$.

The standard approach to solve equations of type \eqref{eq:d_Sproj} is based on the uniform discretization with constant quadratures. The reader can find more sophisticated and efficient procedures, namely, the \emph{semi-analytic} (SA) and \emph{spline-based} (SB) methods, described in Ref.~\citep{Jackura:2020bsk}. We start by introducing the $i\epsilon$ prescription in $\Fc$. This shifts the pole slightly away from the axis of integration. Next, we introduce a discretized mesh in momentum space to numerically approximate the integral equation by a system of $N$ linear equations. We denote the $\epsilon$-dependent and $N$-dependent solution of our equation by explicit $N, \epsilon$ subscripts, i.e., we write $d_{N,\epsilon}(p',p)$. The true ladder amplitude $d(p',p)$ is given in the ordered limit as $N \to \infty$ and then $\epsilon \to 0$. Having a uniform mesh of points $\{k_n\}$, one rewrites Eq.~\eqref{eq:d_Sproj} as,
    \begin{align}
    \label{eq:dNeps_eq}
    d_{N,\epsilon}(p',p) 
    &=
    G(p',p) 
    + \sum^{N-1}_{n=0} \frac{\Delta k_n \, k^2_n }{ (2\pi)^2 \, \omega_{k_n}}
    \, G(p',k_n)
    \, \Fc(k_n) \, 
    d_{N,\epsilon}(k_n,p) \, .
    \end{align}
We consider $d$ as a matrix in the space defined by the set $\{k_n\}$, with matrix elements $d_{{n'}{n}} =d_{N,\epsilon}(k'_{n}, k_{n})$. This allows us to consider Eq.~\eqref{eq:dNeps_eq} as a linear system, with a solution,
    \begin{align}
    \label{eq:Mphib_sol}
    d_{N,\epsilon}(p',p)
    &=
    \left[ M^{-1} \, {G} \right]_{n'n}\bigg|_{k_n'=p',\,k_{n}=p} \ ,
    \end{align}
where $M$ is a matrix given by,
    \begin{align}
    \label{eq:Bphib}
    M_{n' {n}}
    &=
    \delta_{{n'}{n}}
    - \frac{ \Delta k_n \, k_{n}^2 }{ (2\pi)^2 \, \omega_{k_{n}}}
    \, G_{S}(k_{n}', k_{n})
    \, \Fc(k_{n})  \, .
    \end{align}
In the last step, for a large value of $N$, using Eq.~\eqref{eq:dNeps_eq}, one interpolates the solution, Eq.~\eqref{eq:Mphib_sol}, to external momenta $p',p \to q$, which is equivalent to taking the limit defined in Eq.~\eqref{eq:limit}.

    \begin{figure}[b!]
        \centering
        \includegraphics[ width=0.9\textwidth]{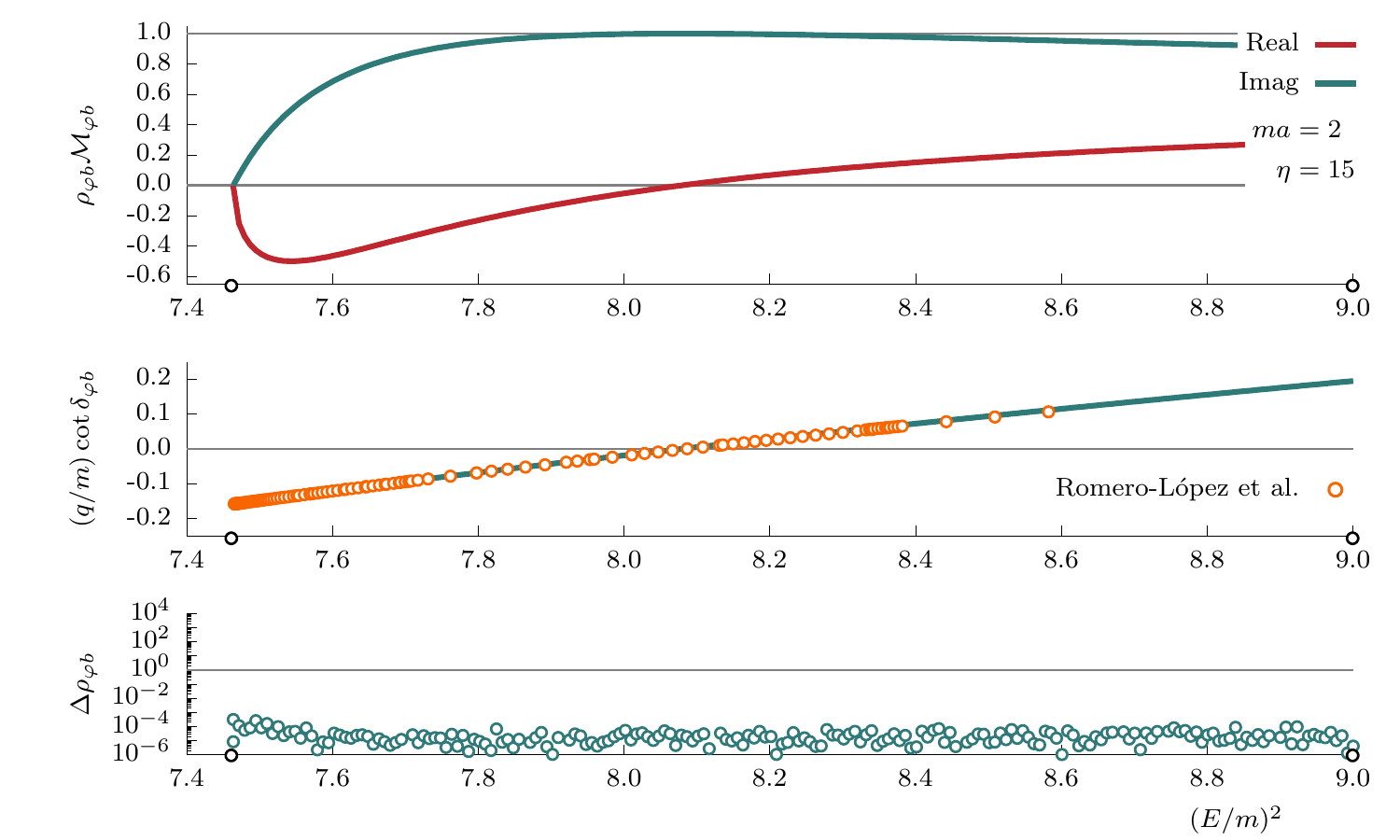}
        \caption{ Solution for the $\varphi b$ scattering amplitude as a function of $s = E^2$ below the three-particle threshold for $ma = 2$. The top panel shows the real (red) and imaginary (blue) parts of $\rho_{\varphi b} \Mc_{\varphi b}$. The open circles on the real axis indicate the $\varphi b$ and $3\varphi$ thresholds. The middle panel shows the resulting $q\cot\delta_{\varphi b}$ (blue), with the open orange points being solutions from the three-particle finite volume formalism taken from Ref.~\cite{Romero-Lopez:2019qrt}. The bottom panel shows the unitarity deviation. 
        }
    \label{fig:a2_avg}
    \end{figure}

In practice, we calculate the amplitude for several values of sufficiently large $N$ and small $\epsilon$ to perform an extrapolation of the numerical result to the $N\to\infty$ and $\epsilon \to 0$ solution. For that, we determined the asymptotic behavior of the error for large $N$ and small $\epsilon$, finding $ d(p',p) = d_{N,\epsilon}( p', p) + \mathcal{O}\left(e^{-\eta}\right)$, where $\eta = 2\pi N \epsilon_q / q_{\mathrm{max}}$ and $\epsilon_q$ is a function of $s$ linearly proportional to $\epsilon$. The $\eta$ parameter relates $\epsilon$ and $N$, implying that for a given $\epsilon_q$ the matrix size $N$ must be large enough, so that $\eta \gg 1$. The extrapolation is executed by fitting the finite $N$ amplitude for a fixed $s$, $a$, and $\eta$ to a function which is a low-order polynomial of $1/N$. In our computations, we used 11 points in the interval $1000 \leq N \leq 6000$ and found systematic errors to be orders of magnitude smaller than the results themselves.

An example solution for the $\varphi b$ scattering amplitude is shown in the top panel of Fig.~\ref{fig:a2_avg}. We consider the case of $ma = 2$, corresponding to a deeply two-body bound state. The corresponding $q\cot\delta_{\varphi b} = 8 \pi \sqrt{s} \,  \Kc_{\varphi b}^{-1}(s)$ is shown in the middle panel of Fig.~\ref{fig:a2_avg}, together with points computed using the finite volume formalism in Ref.~\cite{Romero-Lopez:2019qrt}. We find an excellent agreement with that independent study. The bottom panel shows the \textit{unitarity deviation}, which is one of the employed measures of the systematic error. For each value of $s$ it is given by $\Delta \rho_{\varphi b} = 100 \times \big| 1 +  \mathrm{Im} \big[{\Mc}^{-1}_{\varphi b,N} (s)\big] / \rho_{\varphi b}(s) \big|$. As $\Delta \rho_{\varphi b} \to 0$, the solution better satisfies the unitarity relation. In our calculations for $ma=2$, we find for all energy points except at threshold show sub-percent deviations, indicating that our numerical approximation is satisfactory for the desired precision. In principle, the SA and SB methods, not described here, allow to decrease $\Delta \rho_{\varphi b}$ arbitrarily low for comparable or smaller matrix sizes $N$.

\section{Bound states in the B-matrix formalism}

We turn to the study of the same system in the ``minimal'' $B$-matrix formalism. In the $B$-matrix equation, the intermediate states are on-shell and have physical energies. The choice of the integration cutoff $\sigma_{\text{min}}=4m^2$ affects the behavior of the amplitude below the three-particle threshold since the integration over the intermediate momentum does not cover the physical region available to bound-state--spectator state. The two-particle bound state pole at $\sigma_b < 4m^2$ is outside of the integration limits for $s > s_{\varphi b}$, and the amplitude $\Mc_{\varphi b}$ defined through the limiting procedure, Eq.~\eqref{eq:limit}, has a wrong two-body threshold behavior. To describe bound-state--spectator scattering in the ``minimal'' unitarity formalism, such a state has to be included as an asymptotic scattering state.

We generalize the $B$-matrix parametrization by explicitly incorporating the channels representing bound-state--particle interactions, i.e., we consider the coupled $\2\to\2$, $\3\to\2$, $\2\to\3$, and $\3\to\3$ scattering processes. Similar approach can be found in Refs.~\citep{PhysRev.174.2022,PhysRev.142.1051}. We introduce the $\n\to\m$ amplitudes $\Mc_{mn}$, which are matrix elements of the $T$ matrix describing different reaction channels. Their precise definition is provided in Ref.~\citep{Dawid:2020uhn}. The amplitude $\Mc_{33}$ has both a connected and disconnected parts, while $\Mc_{32}$, $\Mc_{23}$, $\Mc_{22}$ are connected by definition; thus, we write $\Mc_{nm} = \Ac_{nm}$ for both $n,m\neq3$. The three-particle states are described using kinematic variables $(\p\ell m_\ell)$. In this basis, the partial wave projected $\2\to\3$ amplitude is not a matrix but a ``vector'' $\Ac_{32,\p'}$, the partial wave projected $\3\to\2$ amplitude is a transposed ``vector'' $\Ac_{32,\p}$, while the $\2\to\2$ amplitude $\Ac_{22}$ is a ``scalar''. The two-body system of the bound-state and spectator is described by the total invariant mass squared $s$ or the relative momentum $\bm{k}$ between the particles in the CM frame. The angular orientation $\Omega_{\hat{\bm{k}}}$ of the outgoing spectator's momentum in the two-body system is defined with respect to the incoming spectator's momentum, either in the two-body or the three-body state.

\begin{figure}[t!]
    \centering
    \includegraphics[ width=0.65\textwidth]{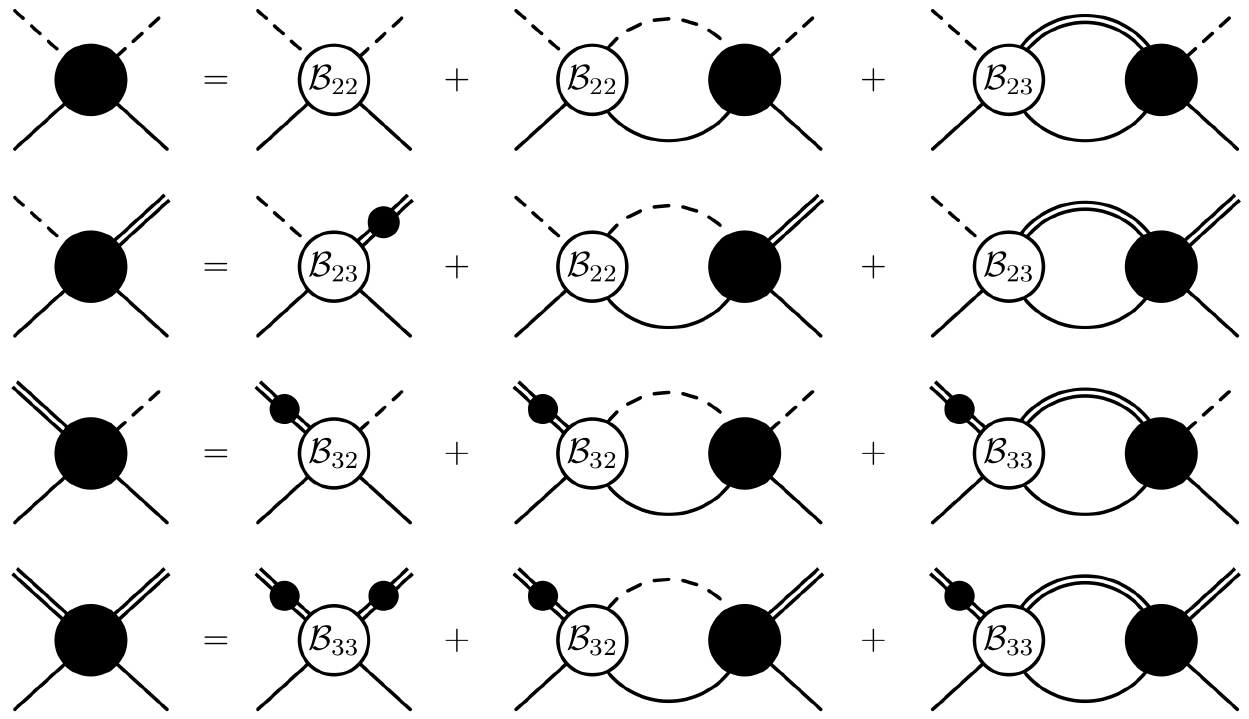}
    \caption{Diagrammatic representation of the multi-channel $B$-matrix framework, Eqs.~\eqref{eq:b-matrix-22}--\eqref{eq:b-matrix-33}. Amplitudes $\Ac_{mn}$ are represented by solid circles and can be differentiated from the different types of external legs. Dashed line represents the two-body bound state. The $\Bc_{33}$ kernel is decomposed as in Fig.~\ref{fig:b-matrix}, while other kernels are real and describe short-range interactions.}
    \label{fig:b-matrix-multi-channel}
\end{figure}

Each new scattering channel has a corresponding $B$-matrix kernel. These are real functions $\Bc_{22} \equiv \Bc_{22}(s,\hat{\bm{k}})$, $\Bc_{23,\p} \equiv \Bc_{23, \ell m_\ell}(s,\sigma_{\p},\hat{\bm{k}})$ and $\Bc_{32,\p'} \equiv \Bc_{32,\ell' m_\ell'}(\sigma_{\p'},s,\hat{\bm{k}})$. Denoting integration over the implicit angular dependence in the intermediate state by $ \int_{\hat{\bm{k}}} = \int \frac{d\Omega_{\hat{\bm{k}}}}{4\pi}$, the generalized $B$-matrix 
parameterization of the connected amplitudes $\Ac_{mn}$ is given by
    \beq
    \label{eq:b-matrix-22}
    \Ac_{22} &=& \Bc_{22} + \int_{\hat{\bm{k}}} \Bc_{22} \, i \rho_{\varphi b} \, \Ac_{22} + \int_{\q} \Bc_{23,\q} \,  \Ac_{32,\q} \, , \\
    \label{eq:b-matrix-23}
    \Ac_{23,\p} &=& \Bc_{23, \p} \, \Fc_{\p} + \int_{\hat{\bm{k}}} \Bc_{22} \, i \rho_{\varphi b} \, \Ac_{23,\p} + \int_{\q} \Bc_{23,\q} \, \Ac_{33,\q \p } \, , \\
    \label{eq:b-matrix-32}
    \Ac_{32,\p'} &=& \Fc_{\p'} \, \Bc_{32,\p'} +  \int_{\hat{\bm{k}}} \Fc_{\p'} \Bc_{32,\p'} \, i \rho_{\varphi b} \, \Ac_{22} + \int_{\q} \Fc_{\p'} \, \Bc_{33,\p'\q} \, \Ac_{32,\q} \, , \\
    \label{eq:b-matrix-33}
    \Ac_{33,\p'\p} &=& \Fc_{\p'} \, \Bc_{33,\p'\p} \, \Fc_{\p} + \int_{\hat{\bm{k}}} \Fc_{\p'} \, \Bc_{32,\p'} \, i \rho_{\varphi b} \, \Ac_{23,\p} + \int_{\q} \Fc_{\p'} \, \Bc_{33,\p'\q} \, \Ac_{33,\q\p} \, .
    \eeq
Their diagrammatic representation is shown in Fig.~\ref{fig:b-matrix-multi-channel} and the formalism can be easily generalized to include other channels. As shown in Ref.~\citep{Dawid:2020uhn}, the amplitudes given above satisfy unitarity above the three-body threshold $s_{3\varphi}$. This representation in general does not satisfy the unitary between the bound-state--particle threshold and the three-body threshold, due to the three-body channel non-zero contribution to the imaginary part of the amplitudes $\Ac_{mn}$ for $m,n<3$.
In an analogy with the $\3\to\3$ formalism, the two-body amplitude $\Fc$ can be amputated in the three-body channels containing pairs, defining $\Ac_{32,\p'} = \Fc_{\p'} \, \widetilde{\Ac}_{32,\p'}$, etc. There is no amputation needed for the bound-state--particle channel, but we write $\Ac_{22} = \widetilde{\Ac}_{22}$ to maintain consistency in the notation. One can write formal solutions to the integral equations given above, using the generalized matrix form,
    \beq
    \label{eq:formal-solution-33}
    \widetilde{\Ac}_{33}
    &=& [\bm{1} -  \Hc_{33} \, \Fc]^{-1} \, \Hc_{33} \, , \\
    \label{eq:formal-solution-22}
    \widetilde{\Ac}_{22} &=& \frac{1}{1- i \rho_{\varphi b} \, \Bc_{22}}  \left[ \Bc_{22} + \Bc_{23} \, \Fc \, [\bm{1} -  \Hc_{33} \, \Fc]^{-1}  \Hc_{32}  \right] \, ,
    \eeq
where the effective kernels are,
    \beq
    \label{eq:modified-kernel-33}
    \Hc_{33,\p'\p} &=& \Bc_{33,\p'\p} + \frac{\Bc_{32,\p'} \, i \rho_{\varphi b} \, \Bc_{23,\p} }{1 - i \rho_{\varphi b} \, \Bc_{22}} \, , 
    \\
    \Hc_{32,\p'} &=& \Bc_{32,\p'} + \frac{\Bc_{32,\p'} \, i \rho_{\varphi b} \, \Bc_{22}}{1-i \rho_{\varphi b} \, \Bc_{22}} \ .
    \eeq
Kernel $\Bc_{33}$, enters the solution for $\wAc_{22}$ through $\Hc_{33}$ and affects the two-body physics as long as $\Bc_{23}$ and $\Bc_{32}$ are nonzero. The $\2\to\2$ scattering occurs even in the absence of the ``direct'' interactions between the bound-state and the spectator, i.e., for $\Bc_{22}=0$. In this case, the dynamics of the two-body scattering is described entirely by the physics involving three-particle interactions.

The analytic properties of the $B$-matrix formalism amplitudes involving the long-range OPE processes were studied in Ref.~\citep{Jackura:2018xnx}. We find that the non-physical analytic properties of the amplitudes appear even in the much simpler case of contact interactions. First, we neglect the effects of long-range interactions by setting the OPE amplitude to zero, $\Gc=0$. We assume that the short-range kernels $\Rc_{33,\p'\p}$ and $\Bc_{23,\p}$, $\Bc_{32,\p'}$, $\Bc_{22}$ are independent on momenta and energies and are rewritten in terms of real coupling constants $\Rc_{33} = g_{33}$,  $\Bc_{23} = g_{23} = \Bc_{32} = g_{32}$, $\Bc_{22} = g_{22}$. In this model, the $B$-matrix equations can be solved exactly, giving,
    \beq
    \label{eq:d33sol}
    \widetilde{a}_{33}(s) &=& \frac{g_{33} + G \, i \rho_{\varphi b}(s) }{1 - g_{22} \, i \rho_{\varphi b}(s) - [g_{33} + G \, i \rho_{\varphi b}(s) ] \, \Ic(s) } \, , 
    \eeq
    \beq
    \label{eq:d22sol}
    \widetilde{a}_{22}(s) &=& \frac{g_{22} + G \, \Ic(s)}{1 - g_{22} \, i \rho_{\varphi b}(s) - [g_{33} + G \, i \rho_{\varphi b}(s) ] \, \Ic(s) } \, .
    \eeq
where $G = g_{32}^2 - g_{33} g_{22}$ and  
    \beq
    \label{eq:integral}
    \Ic(s) = \int\limits_{\sigma_{\text{min}}}^{(\sqrt{s}-m)^2} \frac{d\sigma_k}{2\pi} \, \tau(s,\sigma_k) \, \Fc(\sigma_k) \, .
    \eeq
We write $\widetilde{a}_{nm}$ for the contact-interaction model amplitudes.
Calculating the imaginary part of Eq.~\eqref{eq:d22sol} one obtains, $ \im \widetilde{a}_{22}(s) = \rho_{\varphi b}(s) \, |\widetilde{a}_{22}(s)|^2 + \im  \Ic(s) \, |\widetilde{a}_{32}(s)|^2 $. This agrees with the two-body unitarity for $s \geqslant s_{3\varphi}$, but disagrees for lower energies for non-zero $\im \Ic(s)$. The kernel $\Ic(s)$ governs the analytic structure of the amplitudes. It has the pole at $s=0$ from the $\tau(s,\sigma_k)$ and the right hand cut associated with the three-body threshold $s_{3\varphi}$. In addition, however, nonphysical branch points appear. Instead of the right-hand cut associated with the two-body threshold, $s_{\varphi b}$ there is a left-cut starting at this point. Additionally a left-hand cut appears at $s_{3 \varphi}$ due to the lower integration limit $\sigma_{\text{min}}$ colliding with one of the movable singularities of the integrand.

\begin{figure}[t!]
\begin{center}
\subfigure[~Undispersed model amplitude $\widetilde{a}_{33}(s)$]
{
\includegraphics[width=0.42\textwidth]{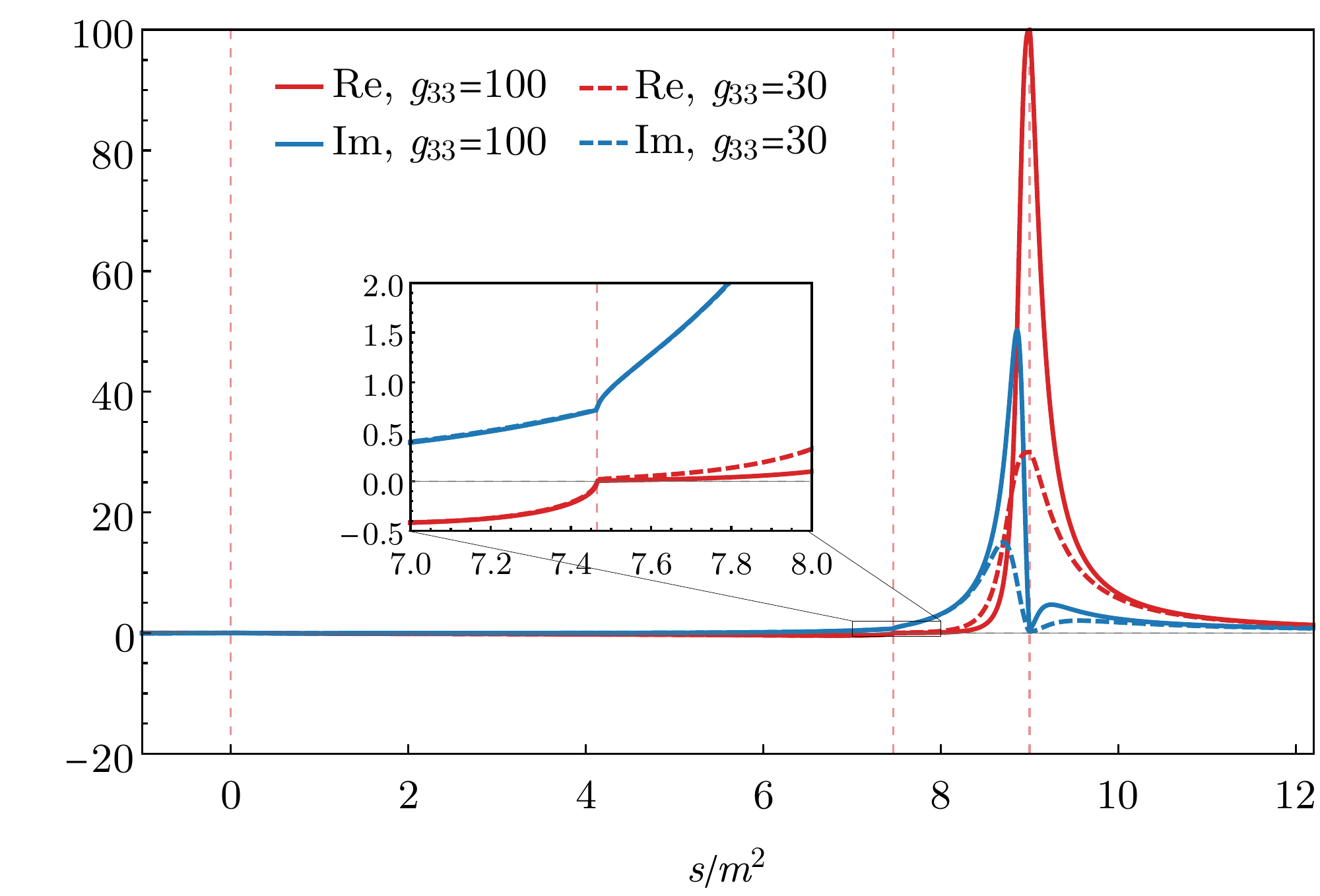}
\label{fig:amp_a}
}
\subfigure[~Dispersed model amplitude $\widetilde{a}_{33,d}(s)$ ]
{
\includegraphics[width=0.42\textwidth]{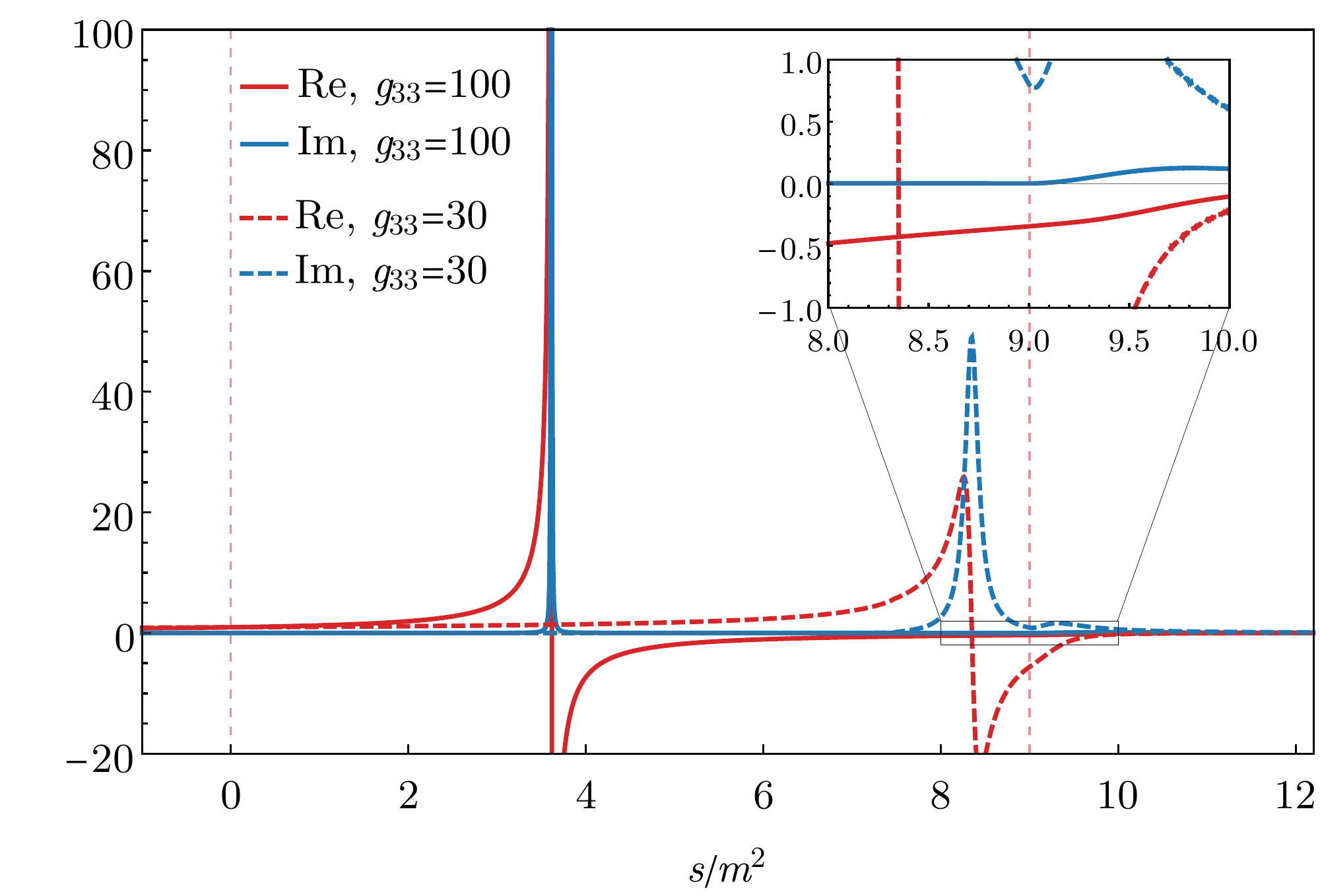}
\label{fig:amp_b}
}
\end{center}
\caption{The comparison between the $\3\to\3$ amplitudes in the model with undispersed (a) and dispersed (b) integral kernel $\Ic(s)$. The value of the couplings are $g_{22}=1$, $G=0$, and $g_{33}=100$ or $g_{33}=30$. The points of non-analyticity are highlighted by the dashed red lines. Amplitude $\widetilde{a}_{33,d}(s)$ has no singularities below $s_{3\varphi}$ other than the three-body bound-state pole at $s_{*}/m^2 \approx 3.618$ for large enough $g_{33}$.}
\label{fig:amp100}
\end{figure}

In consequence, the amplitudes contain non-physical left-hand cuts starting in the vicinity of the physical region and do not satisfy the two-body unitarity relation between the two thresholds. Furthermore, setting $g_{32}=g_{22}=0$, one finds $
\widetilde{a}_{33}(s) = (1/g_{33} - \Ic(s) )^{-1}$. Thus, the location of a three-body bound state is given by conditions $\re \Ic(s_p) = 1/g_{33}$ and $\im \Ic(s_p) = 0$, i.e., it occurs only for a single value of $g_{33}$, since $\im \, \Ic(s)$ vanishes only for one value of energy below the the three-body threshold. Therefore, once the two-body scattering length $a_0$ is fixed, the changes in the three-body coupling $g_{33}$ do not affect the physical predictions of the model. This can be seen in the Fig.~\ref{fig:amp_a}, where the amplitude $\widetilde{a}_{33}$ shows only a nonphysical bump, which scales with $g_{33}$, and a spurious singularity occurring at $s_{\varphi b}$.

The problems with nonphysical singularities can be resolved by a dispersion representation. We construct a new kernel $\Ic_d$ which inherits only the three-body unitarity cut from $\Ic(s)$, 
    \beq
    \label{eq:dispersed-I}
    \Ic_d(s) = \frac{s^2}{\pi}\! \int\limits_{s_{3\varphi}}^\infty \! ds' \frac{\im  \Ic(s')}{(s'-s - i \epsilon)(s' - i\epsilon)^2} \,  ,
    \eeq
The imaginary part of $\Ic_d(s)$ is zero below $s_{3\varphi}$. One also replaces $i \rho_2(s) \to i\rho_{2,d}(s)$ by the Chew-Mandelstam function which removes the nonphysical singularity at $s=0$,
    \beq
    \label{eq:dispersed-rho}
    i \rho_{\varphi b,d}(s) = \frac{s}{\pi} \int\limits_{s_{\varphi b}}^\infty ds' \ \frac{  \rho_{\varphi b}(s')}{s'(s' - s )} \, .
    \eeq
The ``dispersed'' $B$-matrix parametrization results in the $\2\to\2$ amplitude with the proper analytical behavior below $s_{3\varphi}$, $
\widetilde{a}_{22,d}(s) = [\Kc(s)^{-1} - i \rho_{\varphi b,d}(s) ]^{-1}$, where the real $K$ matrix is given by,
    \beq
    \label{eq:KBmatrix}
    \Kc(s) = \frac{g_{22} + G \, \Ic_d(s)}{1-g_{33} \, \Ic_d(s) } \, .
    \eeq
The dispersion relation not only restores the proper analytic properties of the amplitude and ensures its unitarity, but also restores the physical sense of the three-body coupling, see Fig.~\ref{fig:amp_b}.

\section{Conclusions}

In this work, we presented two studies of relativistic bound-state systems in the three-body scattering formalisms. In the first one, we investigated several systematically improvable numerical methods for solving the relativistic three-body on-shell integral equations. Similar equations have been employed in studies involving realistic hadronic systems \citep{Sadasivan:2020syi,Hansen:2020otl}. We tested the validity of our approach by employing several checks on the systematic errors and showed agreement with the calculation presented in Ref.~\cite{Romero-Lopez:2019qrt}, which involved the finite-volume quantization condition. Our methodology can be adopted to resonating systems and processes with non-zero angular momenta. Additionally, the strategies presented here can be used to include the short-distance three-body $K$ matrices, determined from the growing body of the three-body LQCD calculations \citep{Detmold:2008yn, Detmold:2008fn, Mai:2018djl, Mai:2019fba, Horz:2019rrn, Culver:2019vvu, Blanton:2019vdk, Fischer:2020jzp, Alexandru:2020xqf, Mai:2021nul, Hansen:2020otl,Blanton:2021llb}.

Using the same system as the case study we discussed the analytic features of the ``minimal" $B$-matrix formulation. It features spurious left-hand singularities and prevents one from extracting amplitudes involving a two-body bound-state from the three-body ones. We eliminated those shortcomings by including the physical, coupled channels and employing dispersion relations to push the spurious singularities into nonphysical sheets. This led to controllable and correct amplitudes, which satisfy unitarity constraint above all relevant thresholds. After these generalizations, the $B$-matrix formalism can be used to study the coupled-channels problems. However, more general dispersion procedure is required to ensure the analyticity of amplitudes for the cases beyond contact interaction \citep{Jackura:2018xnx}.
\vspace{-9pt}
\section{Acknowledgements}

I would like to thank my collaborators Raúl Briceño, Md Habib E Islam, Andrew Jackura, Connor McCarty, and Adam Szczepaniak for their contributions in completing this work. This work was supported by the U.S. Department of Energy under Grants No.~DE-AC05-06OR23177 and No.~DE-FG02-87ER40365.

\bibliographystyle{JHEP}
\bibliography{main}

\end{document}